\documentclass[aps,prb,twocolumn,superscriptaddress,nofootinbib]{revtex4-2}

\usepackage[utf8]{inputenc}
\usepackage[T1]{fontenc}

\usepackage[version=4]{mhchem}
\usepackage{graphicx}
\usepackage{amsmath,amssymb}
\usepackage{hyperref}
\usepackage{xcolor}
\usepackage{dcolumn}
\usepackage{bm}

\hypersetup{
    colorlinks=true,
    linkcolor=blue,
    citecolor=blue,
    urlcolor=blue
}

\usepackage[version=4]{mhchem}
\usepackage{graphicx}
\usepackage{dcolumn}
\usepackage{bm}

\makeatletter
\def\section#1{\par\vspace{2.2em}
\refstepcounter{section}
\noindent{\large\bfseries \arabic{section}\,|\,#1}\par\vspace{0.8em}}
\makeatother

\makeatletter
\renewcommand{\thesubsection}{\arabic{section}.\arabic{subsection}}

\def\subsection#1{\par\vspace{1.6em}
\refstepcounter{subsection}
\noindent{\normalsize\bfseries \thesubsection\,-\,#1}\par\vspace{0.8em}}
\makeatother



\begin{document}


\preprint{APS/123-QED}

\title{Atomic scale demonstration of ferromagnetism in a single layer \ce{FeCl2} on Au(111)}%

\author{Adriana E. Candia}
\affiliation{Laboratorio de Microscopias Avanzadas, Universidad de Zaragoza, E-50018, Spain}
\affiliation{Centro de F\'isica de Materiales, CSIC/UPV-EHU, 20018 Donostia-San Sebasti\'an, Spain}
\affiliation{Instituto de F\'isica del Litoral (IFIS-Litoral), CONICET-UNL, 3000 Santa Fe, Argentina}

\author{Eliecer Pel\'aez-Sifonte}
\affiliation{Instituto de Nanociencia y Materiales de Arag\'on, CSIC-Universidad de Zaragoza, E-50009, Spain}

\author{Amitayush Jha Thakur}
\affiliation{Centro de F\'isica de Materiales, CSIC/UPV-EHU, 20018 Donostia-San Sebasti\'an, Spain}

\author{Sebastien E. Hadjadj}
\affiliation{Centro de F\'isica de Materiales, CSIC/UPV-EHU, 20018 Donostia-San Sebasti\'an, Spain}

\author{Samuel Kerschbaumer}
\affiliation{Centro de F\'isica de Materiales, CSIC/UPV-EHU, 20018 Donostia-San Sebasti\'an, Spain}

\author{Aymeric Saunot}
\affiliation{Centro de F\'isica de Materiales, CSIC/UPV-EHU, 20018 Donostia-San Sebasti\'an, Spain}

\author{Martina Corso}
\affiliation{Centro de F\'isica de Materiales, CSIC/UPV-EHU, 20018 Donostia-San Sebasti\'an, Spain}

\author{Maxim Ilyn}
\affiliation{Centro de F\'isica de Materiales, CSIC/UPV-EHU, 20018 Donostia-San Sebasti\'an, Spain}

\author{Jorge Lobo-Checa}
\affiliation{Instituto de Nanociencia y Materiales de Arag\'on, CSIC-Universidad de Zaragoza, E-50009, Spain}
\affiliation{Departamento de F\'isica de la Materia Condensada, Universidad de Zaragoza, E-50009, Spain}

\author{Celia Rogero}
\affiliation{Centro de F\'isica de Materiales, CSIC/UPV-EHU, 20018 Donostia-San Sebasti\'an, Spain}

\author{David Serrate}
\email{serrate@unizar.es}
\affiliation{Laboratorio de Microscopias Avanzadas, Universidad de Zaragoza, E-50018, Spain}
\affiliation{Instituto de Nanociencia y Materiales de Arag\'on, CSIC-Universidad de Zaragoza, E-50009, Spain}
\affiliation{Departamento de F\'isica de la Materia Condensada, Universidad de Zaragoza, E-50009, Spain}

\date{\today}

\begin{abstract}
\ce{FeCl2} is a promising single-layer material with sizeable magnetic susceptibility and insulating character that can be easily grown by molecular beam epitaxy on various surfaces. In order to include it into the select palette of van der Waals materials used to engineer functional heterostructures, it is necessary to confirm its magnetic and electronic ground states, and understand the influence of the supporting substrate. In this work, we unambiguously demonstrate ferromagnetic ordering in a single-layer \ce{FeCl2} on Au(111) by means of spin-polarized scanning tunnelling microscopy. The material features a relatively wide insulating gap of 3.3~eV and a strongly spin-polarized conduction band that emerges at 1.5~eV above the Fermi level. Atomic scale defects with triangular shape play a primary role in the electronic gap and spin density distribution. Specifically, in a region of 1.6~nm around each defect, the conduction band is locally suppressed and the tunnelling magneto-conductance is reduced a factor of four. By tracking the spin-dependent tunnelling conductance as a function of the applied magnetic field, we record atomically resolved hysteresis loops, revealing a soft ferromagnetic ground state with pronounced out-of-plane anisotropy and coercive fields in the range of 15–50~mT. 

\begin{description}

\item[Keywords]
Transition metal dihalide | Van der Waals materials | spin-polarized STM | \\ 2D magnetism | structural defects

\end{description}
\end{abstract}

\maketitle

\section{Introduction}
The ability of two-dimensional (2D) materials to maintain long-range magnetic ordering has significant technological implications. They can induce magnetic functionalities, such as spin filtering or spin current generation, when incorporated into van der Waals (vdW) heterostructure stacks~\cite{song_giant_2018}. One primary drawback is the instability of collective magnetic behaviour against thermal fluctuations in the thermodynamic limit~\cite{mermin_absence_1966}.
Due to magneto-crystalline anisotropy~\cite{gong_cr2ge2te6_2017,huang_cri3_2017,lobo-checa_fedca_2024} and finite size effects~\cite{jenkins_breaking_2022}, a magnetic ground state can be stabilized in some compounds. Most of the few examples of magnetic two-dimensional (2D) materials are obtained by mechanical exfoliation. An exception are transition metal (TM) halides, which can be grown by molecular beam epitaxy via stoichiometric powder sublimation. This allows for a much broader range of fabrication strategies and provides greater design flexibility for atomically thin devices. 

In the strict 2D limit, extended ferromagnetic order is known to occur in TM trihalides such as \ce{CrI3}~\cite{huang_cri3_2017} and \ce{CrCl3}~\cite{bedoya_intrinsic_2021}. TM dihalides (TMDH) would be advantageoud because of their simpler stoichiometry (TMX$_{2}$). However, their magnetic properties have not yet been well established. First-principles calculations predict that the dichlorides (X=\ce{Cl}) with 1T phase (TM = Fe, Co, or Ni) are ferromagnetic Mott insulators~\cite{botana_electronic_2019}. Nonetheless, the possibility of competing exchange interactions (direct exchange between the transition metal ions and superexchange through the halogen p-orbitals) and the geometrical frustration inherent to any triangular unit cell can promote other types of magnetic textures, such as antiferromagnetic arrangements or non-collinear ordering~\cite{li_high_2020,bo_magnetic_2024}.

Recent experimental studies have provided growing evidence of magnetic ordering in this class of materials. The most common strategy is to track the dichroic contrast stemming from the TM ions as a function of field and temperature, using either X-ray photoemission spectroscopy~\cite{bikaljevic_noncol_2021,hadjadj_febr2_2023,aguirre_fecl2_2024,kerschbaumer_cocl2_2025} or Raman scattering~\cite{jiang_general_2023}. These are spatial-averaging techniques that cannot address individually each layer when different thicknesses coexist. In such cases, interlayer coupling effects may arise, concealing the intrinsic properties of the single‑layer system. These mesoscopic techniques also fail to address the impact of point defects on the local magnetization, or the presence of non-collinear spin arrangements~\cite{bikaljevic_noncol_2021,hadjadj_febr2_2023} arising from geometrical frustration and competing exchange interactions. Consequently, in this potentially heterogeneous system, the observation of a hysteresis cycle or spontaneous remanence at zero field has been elusive to date. Such observation would serve as positive, unquestionable proof of long-range ferromagnetic ordering, as has been demonstrated for other two-dimensional (2D) ferromagnetic materials~\cite{gong_cr2ge2te6_2017,huang_cri3_2017,lobo-checa_fedca_2024}. 

In this context, local measurement techniques become essential for understanding the magnetic character of TMDHs. The available magnetometry methods offering the ultimate spatial resolution are spin-polarized scanning tunnelling microscopy (SP-STM)~\cite{bode_spin-polarized_2003} and inelastic spectroscopy of a nickelocene (\ce{NiCp2}) molecule adsorbed on the tip apex~\cite{verlhac_NiCp_2019,fetida_Single_2024}. The latter is a technically demanding variant of inelastic spin-flip spectroscopy, and requires computational modelling of the low energy excitations of \ce{NiCp2} using multiple adjustable parameters~\cite{fetida_Single_2024,pinar_nickelocene_2024}. It has been applied to probe out-of-plane spin moments in \ce{FeCl2}~\cite{aguirre_fecl2_2024}. However, similar signals can be retrieved from isolated paramagnetic spins on surfaces~\cite{verlhac_NiCp_2019,song_highly_2024,wackerlin_role_2022} and the spin excitations of the \ce{NiCp2} are insensitive to the sign of the surface magnetic moment. All this renders the \ce{NiCp2}-based inelastic spectroscopy insufficient on its own to uncover collective magnetic ordering.

Here, we resort to SP-STM to investigate the dependence of the spin density of a single \ce{FeCl2} slab on Au(111) as a function of magnetic field. We find that the local magnetization of each \ce{FeCl2} flake corresponds to that of a soft ferromagnet with pronounced out-of-plane magnetocrystalline anisotropy, with nearly squared magnetization loops and coercive fields of a few tens of mT. We also reveal the spectroscopic fingerprints of the tunnelling differential conductance associated to the material's intrinsic spin-polarized bands. The magnetization and density of states related to the conduction band are strongly reduced near the characteristic atomic scale defects of the \ce{FeCl2} monolayer (ML) on Au(111). Furthermore, we investigate a set of distinct resonances appearing around the Fermi level, which could be mistaken for metallic states of \ce{FeCl2}, but are originating from the quasiparticle interference patterns of the shifted surface state of Au(111). This phenomenon is analogous to that observed in \ce{CoCl2}/Au(111)~\cite{kerschbaumer_cocl2_2025} and graphene on Cu(111)~\cite{gonzalez_graphene_2016}.


\begin{figure*}[t]
\centering
\includegraphics[width=1.8\columnwidth,keepaspectratio]{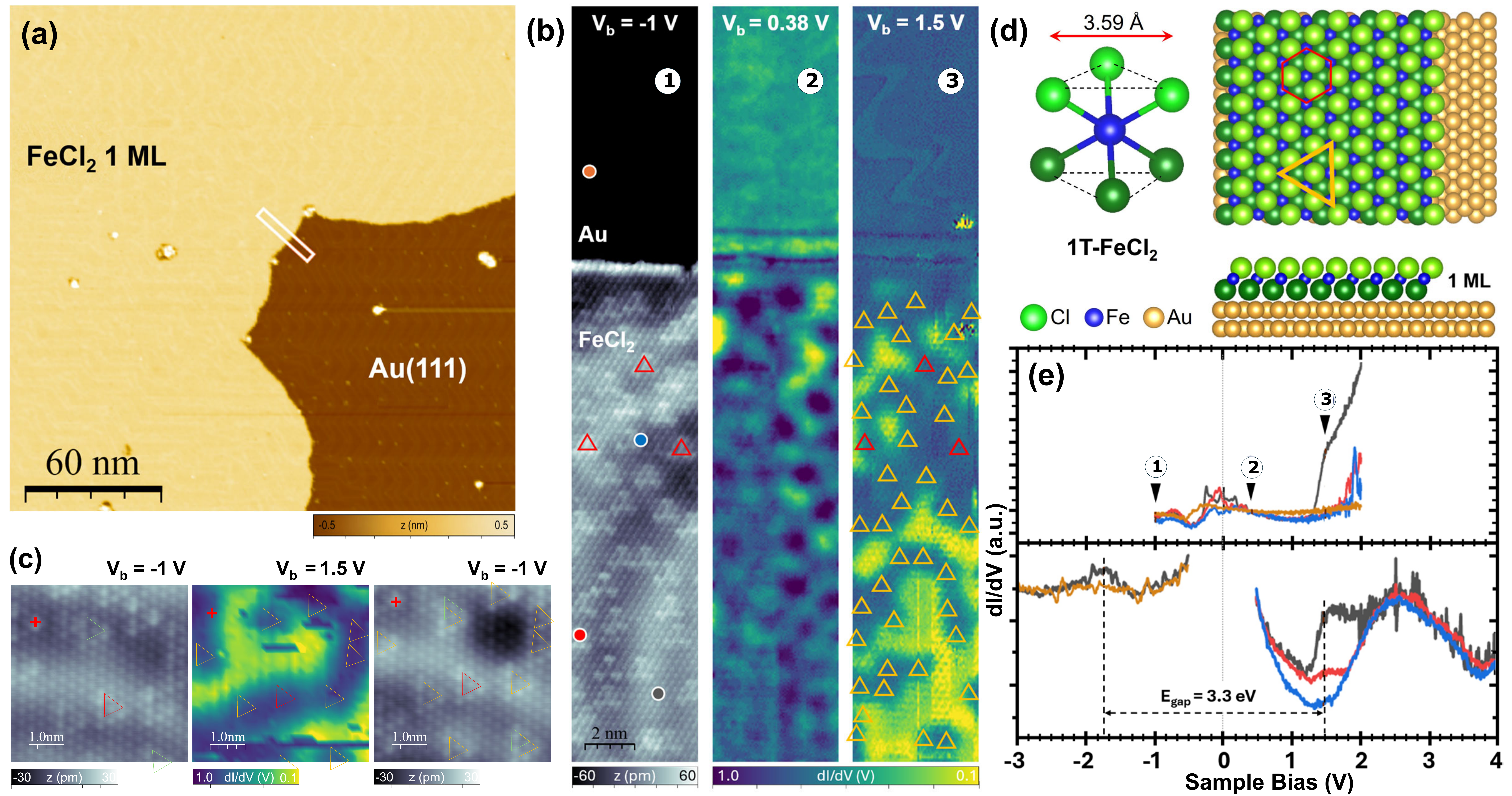}
\caption{\textbf{Correlation of electronic features and structural defects in \ce{FeCl2}}. \textbf{(a)}  Survey of the \ce{FeCl2} ML on Au(111). \textbf{(b)} Atomically resolved topography of the rectangular region shown in (a), together with the $dI/dV$ maps of the same region at 0.38~V and 1.5~V. Triangles mark the position of triangular defects (TD) observed in the topography. Red triangles are a guide pointing to exactly the same position in topography and $dI/dV$ images. \textbf{(c)} Atomically resolved images at $V_b=-1$~V before (left panel) and after (rigth panel) scanning the same area at $V_b=1.5$~V ($dI/dV$ map, middle pannel). The red cross marks the same atomic position in all three images, and the red triangle also serves as a reference. TD are  outlined by green[yellow] triangles before[after] the 1.5~V scan. The two TD singled out in the left panel shift by one lattice site in the right panel. \textbf{(d)} Ball model representation of a ML 1T-\ce{FeCl2}/Au(111), shown in side and top views. The red arrow indicates the unit cell. Each ML is composed of three atomic planes arranged in such a way that the Fe ion sits in an octahedral environment. \textbf{(e)} $dI/dV$ spectroscopy acquired in the positions marked in (b) by dots with the same colour code. Top and bottom panel curves are taken in constant height and constant current modes, respectively. The feature associated to the conduction band (3) is only visible in sample spots without TD nearby. STM parameters: (a) $V_s/I_s=1.3$~V/100~pA; (b) From left to right $V_s/I_s=$-1~V/100~pA, 0.38~V/70~pA, and 1.5~V/150~pA; lock-in modulation $V_{mod}=$10~mV rms. (c) From left to right: $V_s/I_s=$-1~V/150~pA; $V_s/I_s=$-1~V/15~pA and $V_b=1.5$~V with $V_{mod}=10$~mV; $V_s/I_s=$-1~V/150~pA. (e) Constant height spectroscopy in top panel is initially regulated at 1~V/200~pA and $V_{mod}=5$~mV. Regulation parameters for constant current measurements in bottom panel are 1.5~V/70~pA and -1~V/70~pA for the positive and negative bias branches respectively, and $V_{mod}=10$~mV. Temperature is 4.3~K.}
\label{fig_general}
\end{figure*}


\section{Results and discussion}\label{sec2}
Figure~\ref{fig_general}~(a) shows an extended \ce{FeCl2} ML grown by molecular beam epitaxy (see Supporting Fig.~S1). Previous studies show that the \ce{FeCl2} ML crystallizes in the \emph{P$\overline{3}$m1} space group, taking the so called 1T structure. This can be described as one plane of Fe atoms in a hexagonal lattice, sandwiched by two atomic planes of Cl, as shown in the top and side views of Fig.~\ref{fig_general}~(d). The Cl atoms bond to the Fe ions forming an octahedral environment. This gives rise to the splitting of the d-states into the t$_{2g}$ and e$_g$ sub-spaces, which determine the magneto-crystalline anisotropy depending on the electron filling of the TM ion in the high spin state (d$^6$ and $S=2$ for Fe$^{2+}$)~\cite{ashton_two-dimensional_2017}. The Cl atoms in the upper plane form a hexagonal lattice as shown in Figs.~\ref{fig_general}~(b),(c). We obtain a lattice parameter 3.59(5)~\AA\ in good agreement with existing experimental references~\cite{aguirre_fecl2_2024,zhou_evidence_2024}.

\subsection{Influence of structural defects on the electronic structure}
The images in Figs.~\ref{fig_general}~(b)-(c) show some triangular defects (TD) that can only be imaged with atomic resolution at negative bias $V_b<-0.5$~V. Figure~\ref{fig_general}~(b) exemplifies the abundance of TD in the \ce{FeCl2} ML. The length of the TD side equals to three \ce{FeCl2} unit cells, and they display a slightly brighter contour than the surrounding atoms in topography images at $V_b=-1$~V. The yellow triangle in Fig.~\ref{fig_general}~(d) shows a sketch of the atomic sites involved in the TD. Similar defects have been reported in iso-structural vdW materials such as \ce{TiSe2}~\cite{hildebrand_doping_2014} and \ce{HfTe2}~\cite{wang_identification_2023}.

The TDs have a strong impact in the local density of states retrieved by scanning tunnelling spectroscopy. As seen in Fig.~\ref{fig_general}~(e), the differential conductance ($dI/dV$) spectra displays a stepped feature at 1.5~V in distant positions from TD, whereas it is absent on top them. This feature is typical of the onset of a dispersing conduction band (CB), and it has been observed in defect-less \ce{FeCl2} grown on HOPG~\cite{cai_fecl2_2020} and \ce{Bi2Se3}~\cite{klimovskikh_emergence_2025}. The bottom panel of Fig.~\ref{fig_general}~(e) shows $dI/dV$ acquired in constant current mode in a large energy window, where two additional ubiquitous peaks at -1.8~V and 2.6~V are visible, in line with earlier spectroscopy studies~\cite{aguirre_fecl2_2024}. The peak in the occupied states region marks the energy of the valence band (VB). Therefore, the energy gap of \ce{FeCl2} ML amounts to 3.3~eV. In addition, there is a significant intensity around the Fermi level, starting at about 0.2~eV above onset of the Au(111) surface state, which will be discussed in the following section.

The coloured triangles in Figs.~\ref{fig_general}~(b),(c) indicate the positions of the TD. The simultaneous $dI/dV$ maps taken at 1.5~V reveal the the suppression of the CB contribution around the TD. We often observe a tip-induced displacement of the TD, which can be appreciated in Fig.~\ref{fig_general}~(c) in a straight comparison of the topographies before and after measuring the $dI/dV$ map. The map becomes glitchy because of the rapid hopping of the TD during the measurement. We observed that the hopping probability of the TD increases for increasing tunnelling current or bias, with a threshold very close to precisely 1.5~V. Contrary to the behaviour of the CB, the maps of the states near Fermi level (Fig.~\ref{fig_general}~(b) middle panel) do not show any sensitivity to the defects of the \ce{FeCl2}, with an apparently arbitary spatial distribution.

We note that TDs share some distinct characteristics of the mobile polarons found in \ce{CoCl2} and \ce{FeCl2} on HOPG~\cite{liu_atomic-scale_2023,cai_manipulating_2023}: (i) hopping is triggered by injecting electrons with potential energy above the CB onset, and they are attracted by an electric field pointing from sample to tip ($V_b>0$); (ii) the CB is definitely disrupted by them. On the other hand, they cannot be individually created or annihilated on Au(111), as occurs on HOPG~\cite{cai_manipulating_2023}, and there are no ring shaped features in the $dI/dV$ maps~\cite{liu_atomic-scale_2023}. Therefore, a possible explanation to the mobile substitutional defects would be the displacement of a virtual polaron with life time of trapped charge much shorter than the tunnelling rate of electrons, ascribed to the stronger interaction with the Au(111) metallic states, as previously suggested for \ce{CoCl2}/Au(111)~\cite{liu_atomic-scale_2023}. However, unravelling the nature of the TD requires dedicated statistical measurements and theoretical analyses that account for role of the substrate, which lies beyond the scope of this work.

\subsection{Interface state}


\begin{figure*}[t]
\centering
\includegraphics[width=1.8\columnwidth,keepaspectratio]{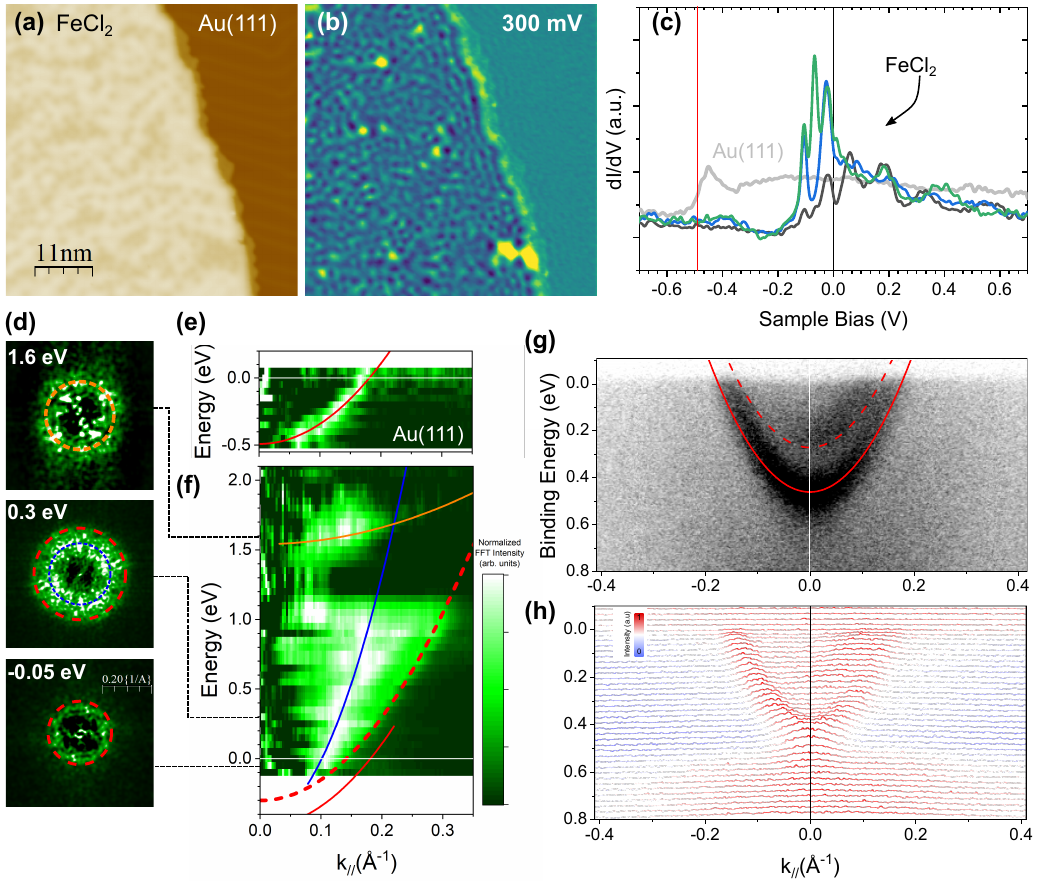}
\caption{\textbf{Electronic structure of \ce{FeCl2} in reciprocal space. (a,b)} Simultaneous topography image and $dI/dV$ map of the same region encompassing an island of \ce{FeCl2} ML and bare Au(111). \textbf{(c)} High resolution $dI/dV$ point spectroscopy of \ce{FeCl2} and reference spectrum of Au(111) (grey). \textbf{(d)} Fast Fourier Transform (FFT) images~\cite{horcas_wsxm_2007} of the \ce{FeCl2} region from $dI/dV$ maps at selected energies. The mathematical processing and raw FFT images can be found in Supplementary Fig.~S2. \textbf{(e,f)} Colour maps of the radial averaged intensities of FFT images as those in (d) as function of energy for the Au(111) and the \ce{FeCl2} regions respectively. The red solid line is a parabolic fit to the experimental points with maximum intensity of the Au(111) radial averages, yielding an effective mass of $m^*=0.23m_e$~\cite{reinert_direct_2001} and energy onset of -0.49~eV (see Supplementary Note~II), the latter fixed to the value marked by the red line in the Au(111) reference spectrum of (c). The dashed red line is constructed with the same $m^*$ and a positive shift of the energy onset up to -0.30~eV. Blue and orange lines are guides to the eye with also parabolic functional form. Circles in (d) have the exact same radius as the $k$ value of the parabolic guides with the same colour code at the indicated energies. \textbf{(g)} ARPES data of a sample with sub-monolayer coverage of \ce{FeCl2} obtained at $T=16$~K and a photon energy $h\nu=17$~eV, overlaid with the same parabolic functions as in (e,f) shifted upwards by 30~meV. \textbf{(h)} Waterfall plot of the inverted ARPES intensity highlighting the two detected dispersing features. STM parameters: regulation current $I_s=60$~pA and $V_{\mathrm{mod}}=30$~mV in panels (a,b) and (d,e,f); $V_s=-1$~V in topography panel (a); spectroscopy in (c) is taken with $V_s/I_s=0.5$~V/100~pA and $V_{\mathrm{mod}}=2$~mV. STM temperature is 4.3~K.}
\label{fig_interface}
\end{figure*}

Before addressing the magnetic response, and in order to confirm the insulating character of this material, it is crucial to clarify the origin of the states crossing the Fermi level.  Their presence makes it tempting to assign metallic character to the \ce{FeCl2} ML. Figure~\ref{fig_interface}~(c) shows selected spectra taken in different \ce{FeCl2} positions with energy resolution of $\sim$1~meV. A set of pronounced peaks with strong spatial dependence emerge, reminiscent of a quantum confinement phenomenon. To investigate the origin of these states we have followed the same procedure as in the case of \ce{CoCl2}~\cite{kerschbaumer_cocl2_2025}. We take constant current $dI/dV$ maps at varying $V_b$ in a region enclosing \ce{FeCl2} and some bare Au(111), as shown in Figs. \ref{fig_interface}~(a-b). Fast Fourier Transformed (FFT) images of the maps (see Fig.~\ref{fig_interface}~(d)) display characteristic quasiparticle interference rings and disks, whose radius equals the maximum scattering vector, $q=2k_\parallel$, with $k_\parallel(E)$ being the momentum of the electrons with potential energy $E$. Figures~\ref{fig_interface}~(e) and~(f) show stack plots of radial averaged profiles of the FFT images as a function of energy ($E=eV_b$), for the Au(111) and the \ce{FeCl2} regions respectively.

The FFT analysis reveals that the \ce{FeCl2} low energy states ($-0.1<V_b<1.2$~V) follow exactly the same parabolic dispersion relation as the Au(111) surface state, with identical effective mass, but a different energy onset of -0.3~eV (see Supplementary Note~II). This is confirmed mesoscopically by angle-resolved photoemission (ARPES) experiments on a sub-ML coverage \ce{FeCl2}/Au(111) sample (Figs.~\ref{fig_interface}~(g-f)). 

The photoemission signal is dominated by a parabolic band that corresponds to the surface state of the bare Au(111), with the same curvature of the dispersion relation deduced from the STM measurements (red solid line) and a 30~meV shift towards
the Fermi level. This slight shift is attributed to lateral confinement effects appearing when the pristine surfaces are partially covered by thin film islands, as reported for (111) Schockley states~\cite{lobo-checa_band_2009}. Notably, we also find a weaker parabolic replica in ARPES maps up-shifted towards the Fermi energy that must originate from the \ce{FeCl2} covered regions. This feature coincides with the dispersion relation obtained in the \ce{FeCl2} islands (red dashed line in Figs.~\ref{fig_interface}~(f,g)). Therefore, we assign the highly structured $dI/dV$ signal around the Fermi level (Fig.~\ref{fig_interface}~(c)) to a modified Au(111) surface state existing at the interface between the substrate and the \ce{FeCl2}, as has been observed for graphene on Cu(111)~\cite{gonzalez_graphene_2016}. The increase in the onset energy of the surface state is caused by the electric field at the interface created by the electric dipoles of the \ce{FeCl2}, which modifies the crystal electric potential at the termination of Au(111) crystal.

Additionally, a second dispersing feature marked with a blue line emerges in the energy window included in Fig.~\ref{fig_interface}~(f). In the FFT images (Fig.~\ref{fig_interface}~(d), middle panel) it manifests more as a ring than a disk. A free electron picture interpretation of this feature leads to a parabolic dispersion relation with much lower effective mass than the interface state. Such high-dispersive features have been reported for Au(111) and assigned to contributions of the bulk electrons~\cite{Schouteden_PRB2009}.

Overall, this analysis establishes that the states near the Fermi level are not intrinsic to \ce{FeCl2} and therefore neither contribute to its magnetic response nor sustain metallic behaviour.

\subsection{Spin polarization of the conduction band}

\begin{figure*}[t]
\centering
\includegraphics[width=1.8\columnwidth,keepaspectratio]{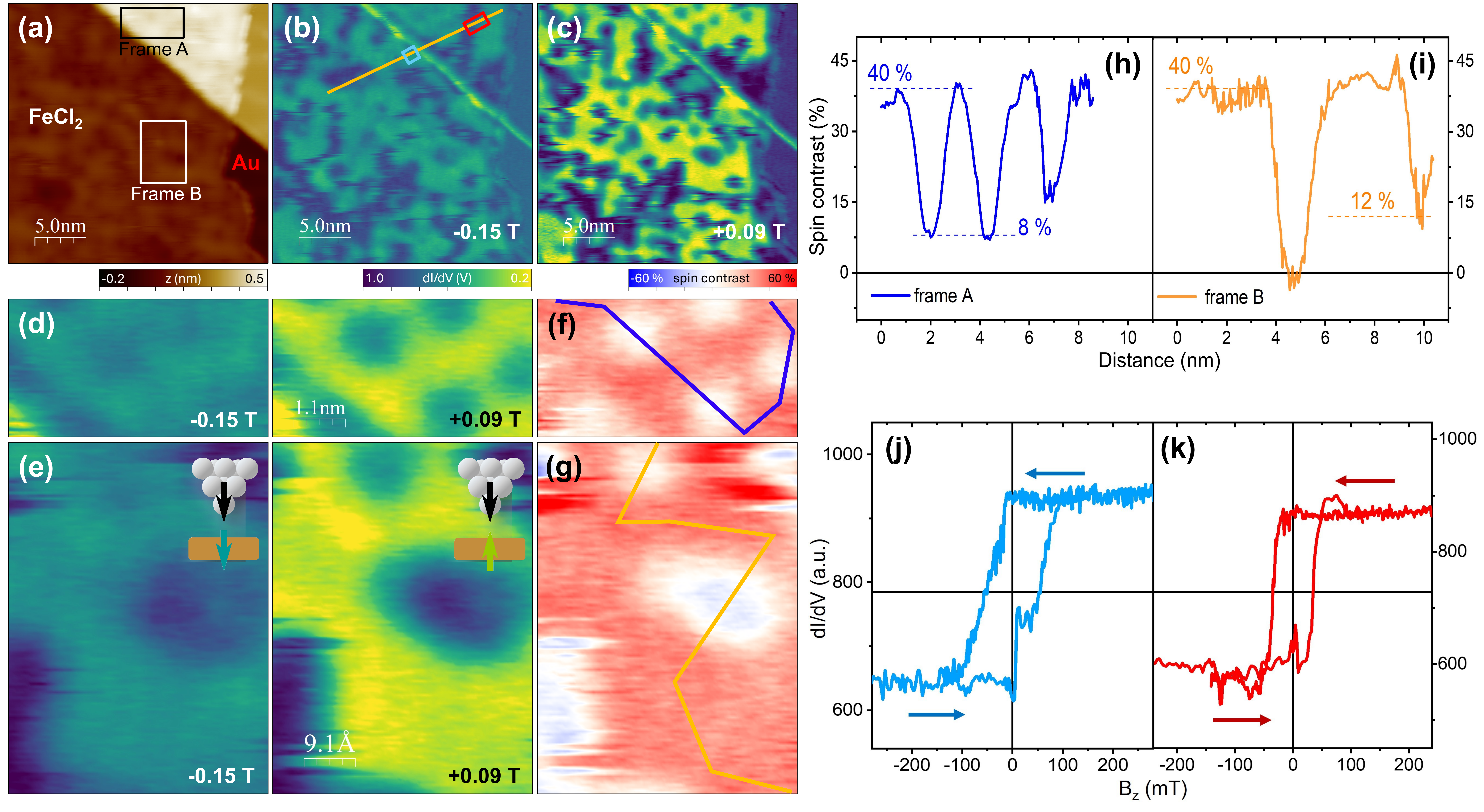}
\caption{\textbf{Spin polarized STM of \ce{FeCl2}. (a)} Topography of a ML \ce{FeCl2} over two different Au(111) terraces. \textbf{(b,c)} $dI/dV$ maps taken after ramping down the field from -2.7~T to -0.15~T ($dI/dV_{\downarrow\downarrow}$, parallel tip and sample magnetization) and further to 0.09~T ($dI/dV_{\uparrow\downarrow}$, antiparallel) after crossing zero. Dark spots occur at the position of the triangular defects (TD). See Supplemental Fig.~S3 for further details on the magnetization loops. \textbf{(d,e)} Zooms in of frames A and B marked in (a) that enclose regions without displacement events of TD (d) during the time elapsed between both images, and with minimal displacement (e). \textbf{(f,g)} Spin contrast of the $dI/dV$ (magneto-conductance) of regions A and B calculated as $\left(dI/dV_{\uparrow\downarrow}-dI/dV_{\downarrow\downarrow}\right)/\left(\langle dI/dV_{\uparrow\downarrow}\rangle +\langle dI/dV_{\downarrow\downarrow}\rangle\right)$, where $\langle dI/dV \rangle$ is the mean differential conductance in the regions without triangular defects. \textbf{(h,i)} Spin contrast profiles outlined in (f,g) by the corresponding coloured lines. \textbf{(j,k)} Magnetic field dependence of the $dI/dV$ contrast for a fixed tip sensitivity direction, thereby representing the local magnetization of the sample. The signal is retrieved from the blue and red areas marked by rectangles in (b), avoiding the influence of TD tip induced displacements. Note that the colour scale for each type of data has the same range throughout the panels. STM parameters: $V_s/I_s=$ 1.5~V/100~pA, lock-in modulation 10 mV rms, $T=1.1$~K.}
\label{fig_spstm}
\end{figure*}

\begin{figure}[b]
\centering
\includegraphics[width=\columnwidth,keepaspectratio]{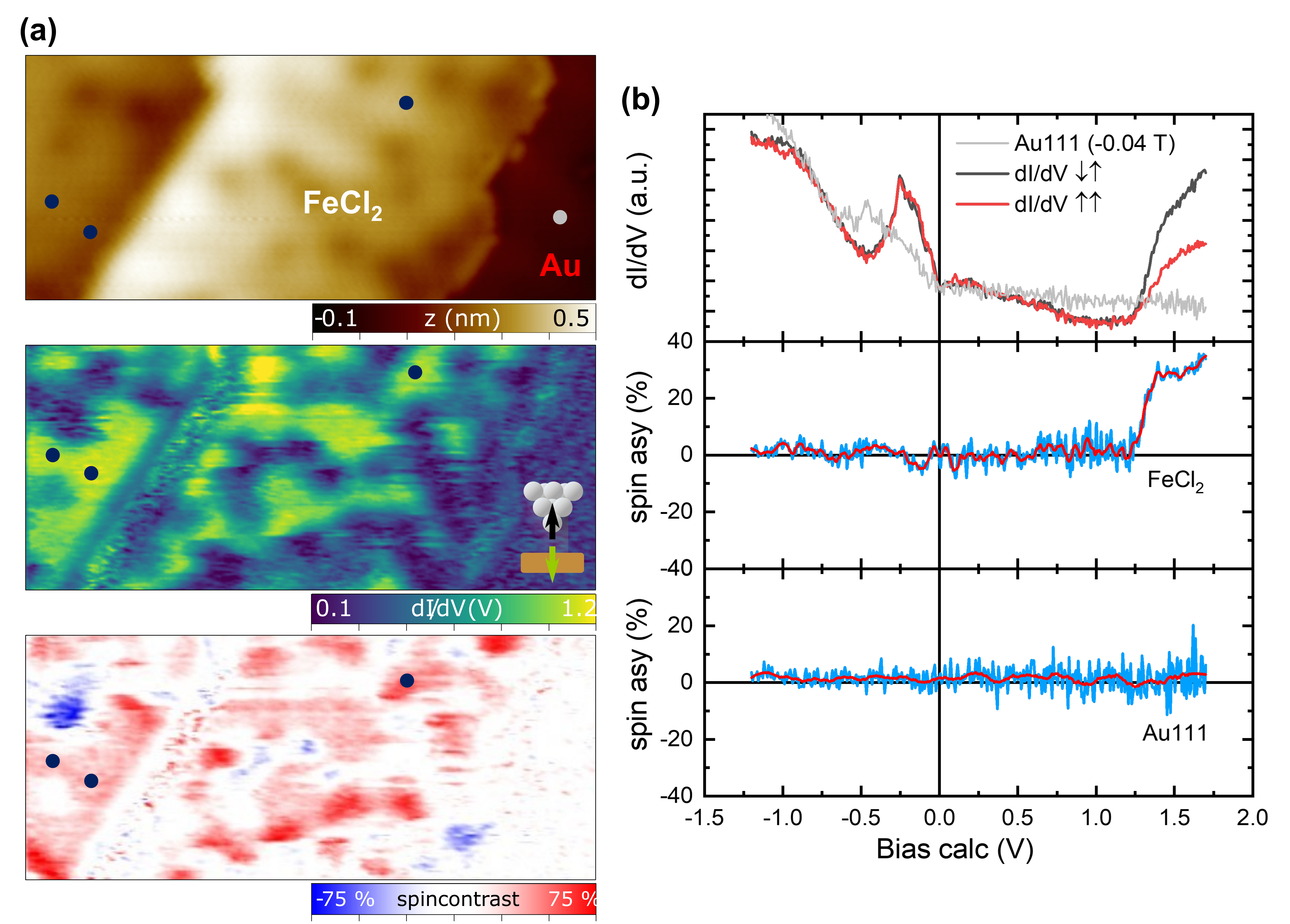}
\caption{\textbf{Spin asymmetry as a function of energy. (a)} Topography of an \ce{FeCl2} island (top); corresponding map for antiparallel tip and sample magnetizations at -0.04~T ($dI/dV_{\downarrow\uparrow}$) (middle); and the spin contrast image (bottom) calculated by subtracting the $dI/dV_{\downarrow\uparrow}$ image above from the the $dI/dV_{\uparrow\uparrow}$ at 2.0~T. \textbf{(b)} Average $dI/dV$ spectra taken at the positions marked by dark blue dots in (a), free from the influence of the TD fluctuations, for the two opposite sample magnetization directions: black spectrum corresponds to the case antiparallel to the tip ($\downarrow\uparrow$), red spectrum for the parallel case ($\uparrow\uparrow$). The middle panel shows the spin asymmetry derived from them (see main text), whereas the bottom panel shows the same quantity for the spectra taken on bare Au, which oscillates around zero reflecting the uncertainty level of this technique. Error bars are given in light blue. STM parameters: $V_s/I_s=$1.5~V/15~pA for all images and spectroscopy curves. Lock-in modulation 8~mV~rms. $T=1.1$~K.}
\label{fig_spsts}
\end{figure}

Now we turn our attention to the central result of this work: the direct and quantitative measurement of the magnetic character of \ce{FeCl2}. The high energy window of Fig.~\ref{fig_interface}~(f) displays the FFT analysis of the $dI/dV$ interference maps of the CB ($V_b>1.5$~V). Figure~\ref{fig_interface}~(d) (upper panel) shows the interference pattern derived from the $dI/dV$ map at $V_b=1.6$~V, exhibiting again a characteristic interference ring. This ring's radius is energy dependent, as evidenced from the additional curved feature developing at exactly 1.5~V, and highlighted by an orange line in Fig.~\ref{fig_interface}~(f). This feature is compatible with a conduction band much flatter than the low energy interface state. 

Spin polarized STM experiments on the \ce{FeCl2} CB at 1.5~eV have been conducted with an out-of-plane sensitive bulk Cr-tip (see Experimental methods). The results are summarized in Fig.~\ref{fig_spstm}. The topography image (Fig.~\ref{fig_spstm}~(a)) and the $dI/dV$ map (Fig.~\ref{fig_spstm}~(b)), taken under an axial field of $B=-2.7$~T, display the same fuzziness observed in Fig.~\ref{fig_general}~(c), which is a consequence of the rapid jumping of the TD when tunnelling to the CB (see Supplementary Note III). Therefore, genuine variations in the magnetic‑field‑dependent contrast can only be reliably probed in regions where no TD crosses the area during the acquisition of the spin‑resolved images. At a large field of $B=-2.7$~T, the sample and the tip are magnetized in the same direction (see magnetic history of this experiment in Supplementary Fig.~S3). Figure~\ref{fig_spstm}~(c) shows the $dI/dV$ map after reversing the sample magnetization at $B=0.09$~T, as illustrated by the sketches of tip and sample magnetic moments in (e). With the exception of the regions where mobile TDs disrupt the contrast, it is clear that the tunnelling conductance is greatly enhanced when tip and sample magnetization are antiparallel. Without loss of generality, we choose the sample's majority spin oriented towards the surface normal in this magnetic configuration.

Figures~\ref{fig_spstm}~(d)-(e) show selected \ce{FeCl2} regions without significant influence of TDs displacement, together with a schematic representation of the tip and sample magnetization in each case. The difference images (Fig.~\ref{fig_spstm}~(f)-(g)) between the positive and negative sample magnetization represent a map of the local spin density (see Supplementary Fig.~S3~(c) for a spin density map of the entire region of interest). It is worth noticing that the dark regions at $V_b=1.5$~V are centered at each TD (Figs.~\ref{fig_general}~(b-c)), and coincide with the regions where the spin density maps have local minima. In particular, as shown in the profiles of Fig.~\ref{fig_spstm}~(h)-(i), the maximum magnetic contrast in the defect free regions amounts to 40\% of the average conductance, whereas at the positions of the TD the value drops significantly to $\sim$10\%. This is still above the noise level that we retrieve from the bare Au(111) regions, which has a value of $\sim$2\% under our experimental conditions (see Supplementary Fig.~S3~(e)). The TDs therefore suppress the spin density around them, but do not fully cancel it out. 

The field dependent $dI/dV$ conductance is plotted in Fig.~\ref{fig_spstm}~(j-k) for the sample spots marked in Fig.~\ref{fig_spstm}~(b). The spin resolved measurement captures a clear squared magnetization loop, characteristic of a soft ferromagnet with out-of-plane easy axis. The coercive field value oscillates between 15 and 50~mT. Such inhomogeneity throughout a \ce{FeCl2} island can be attributed to domain wall pinning and inhomogeneous exchange constants in the vicinity of the TD.

Finally, Fig.~\ref{fig_spsts} illustrates the energy dependence of the spin polarization in \ce{FeCl2}. The top panel in Fig.~\ref{fig_spsts}~(b) shows the average $dI/dV$ of several points far from TD marked in Fig.~\ref{fig_spsts}~(a) for antiparallel ($\uparrow\downarrow$) and parallel ($\downarrow\downarrow$) sample and tip magnetizations. Then the spin asymmetry is calculated as $S_a=\left(dI/dV_{\uparrow\downarrow}-dI/dV_{\downarrow\downarrow}\right)/\left(dI/dV_{\uparrow\downarrow}+dI/dV_{\downarrow\downarrow}\right)$. This quantity is proportional to the sample's spin polarization. It can be seen that the conduction band ($V_b>1.5$~V) of \ce{FeCl2} has a strong spin polarization, whereas the in-gap region is essentially non-spin polarized. The interface states near the Fermi level show a marginal spin asymmetry of $\pm4$\% which is just slightly above the experimental standard deviation (blue error bars).

\section{Conclusions}\label{sec3}

By means of SP-STM we demonstrate that \ce{FeCl2} ML on Au(111) is a soft ferromagnet with an insulating band gap of 3.3~eV and large spin polarization of the conduction band. Triangular defects centred at the position of the bottom Cl atoms interrupt locally the conduction band and induce a decrease of the spin density around them. The domain wall displacement in this material is sufficiently slow as to allow us to register different switching events in regions that are just 10~nm apart. Notably, the magnetization loops of individual \ce{FeCl2} islands display a marked out-of-plane anisotropy character. 

Our results establish the \ce{FeCl2} monolayer as an atomically thin ferromagnetic insulator, and suggest that defect manipulation can be exploited in order to control spin textures and magnetic switching at the nanoscale. This works demonstrates that SP-STM is a highly suitable technique to investigate the magnetic ordering of TM halides, capable of probing the site, energy, aggregation phase and thickness specific spin density. 

\section{Experimental Methods}\label{sec4}
\ce{FeCl2} is deposited from a commercial Kentax resistive evaporator at a temperature of 350~$^\circ$C to achieve a coverage of approximately 0.7~ML. The Au(111) single crystal is previously cleaned by sputtering and annealing cycles at 530~$^\circ$C. To achieve the growth of the 1T crystalline phase, an optimum substrate temperature of 150 $^\circ$C is required. Deviations from these parameters of dose and sample temperature result in the appearance of the well-known precursor phase of the halides~\cite{hadjadj_febr2_2023,kerschbaumer_coverage_2025}. This phase consist in a self assembly of \ce{FeCl2} molecules with completely different electronic and magnetic properties from the 1T monolayer slab. See Supplementary Fig.~S1 for the effect of variations in substrate temperature and coverage. 

All STM measurements have been performed at the SPECS-JT-STM and the Omicron LT-qPlus of the Laboratory for Advanced Microscopy (University of Zaragoza). The whole facility operates under ultra-high-vacuum conditions ($P<1\times10^{-10}$~mbar). The tip is grounded and the tunnelling bias $V_b$ is applied to the sample. Differential tunnelling conductance $dI/dV$ is acquired using a lock-in amplifier at a frequency of 817 or 971~Hz and r.m.s. modulation given by $V_{\textrm{mod}}$. STM images and $dI/dV$ maps were taken either in constant height or in constant current mode, using a regulation distance determined by the set point ($I_s$, $V_s$) indicated at the corresponding caption for each data set. Image processing is performed with WSxM software package~\cite{horcas_wsxm_2007}.

Spin averaged STM experiments are performed with electrochemically etched W tips. For spin resolved measurements, Cr-bulk tips are tips are prepared by electrochemical etching of 1 mm wide pure Cr rods. Afterwards they are field emitted at the STM head (120~V, 1~$\mu$A, 30~min). Before proceeding to the target sample, we confirm that out-of-plane magnetic contrast is readily achieved by positive voltage pulses using the reference signal of 2 epitaxial layers of Fe on W(110)~\cite{pietzsch_observation_2001}. We avoid tip-sample indentation during the whole process, and discard tips with unstable tunnelling gaps. 

ARPES measurements were carried out at LOREA beamline (ALBA Synchrotron) with a base pressure of $1\times10^{-10}$~mbar. The energy resolution (at $h\nu=13$~eV) was $\sim$15~meV and the angular resolution was $\sim$0.1$^\circ$ and $\sim$0.3$^\circ$ along the analyzer slit and perpendicular to it, respectively. Spectra were taken with negative circular polarisation light.

\vspace{1.5em}
\noindent{\bfseries Author Contributions}
\vspace{0.5em}

A.C. and D.S. performed the spin-polarized measurements. Tunnelling spectroscopy was carried out by A.C., D.S. and E.P.S. Growth optimization was performed by E.P.S., and J.L.C.. A.J.T., S.E.H., S.K., A.S., M.C. and M.I. conducted the quasiparticle interference mapping experiments and the ARPES characterization. The project was conceived by D.S. and supervised by C.R. and D.S. The manuscript was written by D.S., A.C., J.L.C. and S.E.H. All authors contributed to the revision and discussion of the final version of the manuscript. \\

\vspace{1.5em}
\noindent{\bfseries Acknowledgments}
\vspace{0.5em}

The authors gratefully acknowledge financial support from the Spanish MICIU/AEI/10.13039/501100011033 and by “ERDF A way of making Europe” through grants PID2022-138750NB-C21 (D.S., J.L.C.), PID2022-138750NB-C22 (M.I., C.R.) and PID2022-140845OB-C65 (M.C.) and PID2023-148225NB-C31 (C.R.), as well as from the Excellence Program Severo Ochoa CEX2023-001286-S (D.S., J.L.C.) and by the IKUR Strategy (HPC-IA and QT 2023-2025). This work has also been supported by the Aragon Government (grants E13-23R and E12-23R) and by ANPCyT-CONICET (grants PICT-INVI-00863 and PICT-2019-04545).

\vspace{1.5em}
\noindent{\bfseries Conflicts of Interest}
\vspace{0.5em}

The authors declare no conflicts of interest.

\vspace{1.5em}

\bibliography{spstm}

\clearpage
\onecolumngrid

\clearpage

\onecolumngrid

\setcounter{section}{0}
\renewcommand{\thesection}{S\arabic{section}}

\begin{center}
{\Large\bfseries Supporting Information}
\end{center}

\setcounter{figure}{0}
\setcounter{table}{0}
\setcounter{equation}{0}

\renewcommand{\thefigure}{S\arabic{figure}}
\renewcommand{\thetable}{S\arabic{table}}
\renewcommand{\theequation}{S\arabic{equation}}

\renewcommand{\figurename}{SUPPLEMENTARY FIGURE}

\vspace{1em}

\begin{center}

{\Large\bfseries
-- Supplementary information for --\\[0.4em]
Atomic scale demonstration of ferromagnetism in a single layer \ce{FeCl2} on Au(111)
\par}

\end{center}

\vspace{1.5em}


\section{Extended Data on sample preparation}
\begin{figure}[h]
\includegraphics[width=\columnwidth,keepaspectratio]{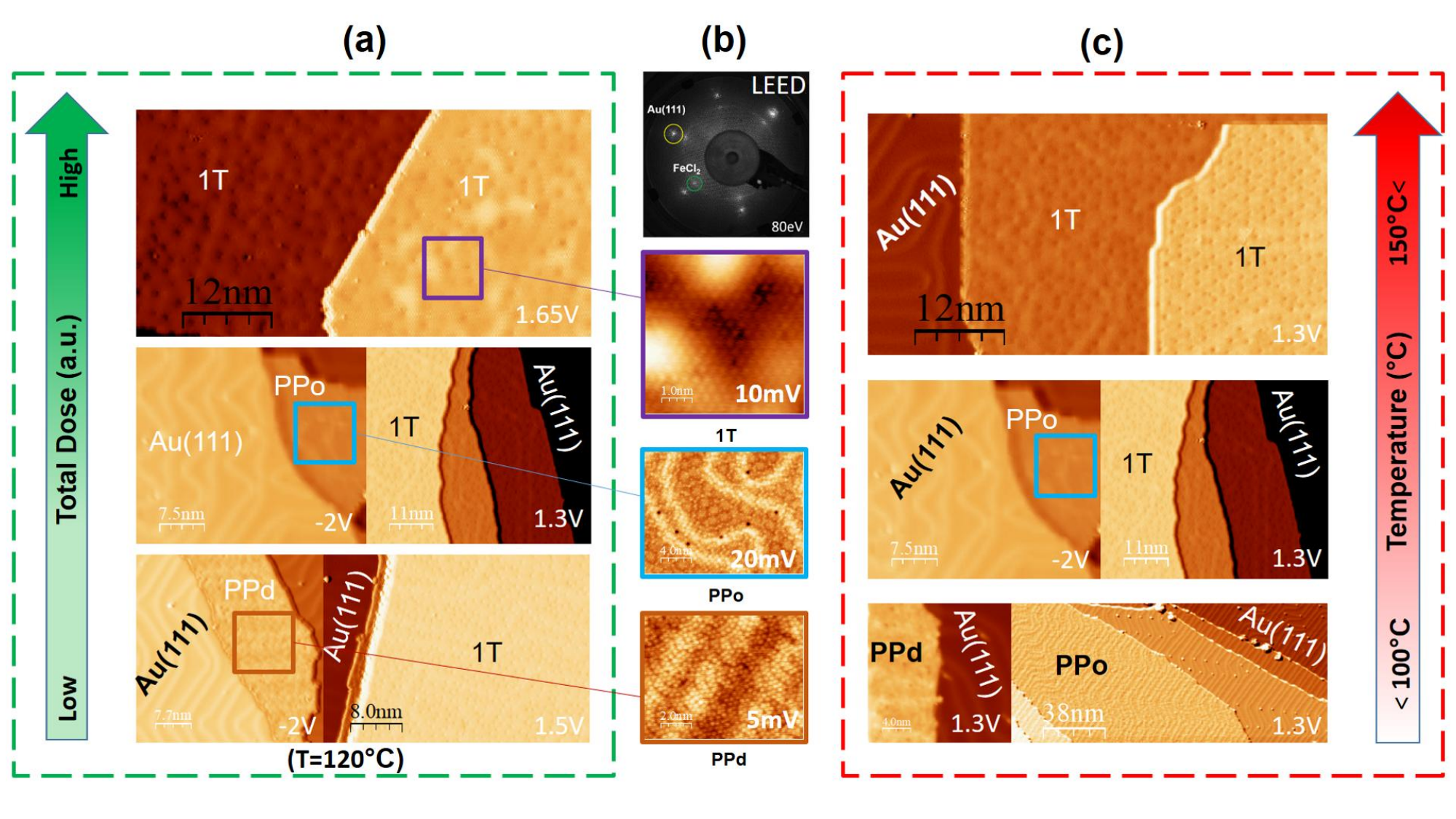}
\caption{\label{fig:FeCl2prep} \small \textbf{(a)} STM topography images of different samples in which the substrate is held at 120$^\circ$C during the deposition of \ce{FeCl2}. The \ce{FeCl2} dose increases in the direction indicated by the left arrow. The different phases identified in this material can be observed. \textbf{(b)} The top panel shows the experimental LEED pattern obtained for \ce{FeCl2} on Au(111)($E=80$~eV). The coloured rectangles show atomically resolved topographies of the different phases of \ce{FeCl2} identified in the samples shown in (a) and (c): Crystalline 1T phase (\textbf{1T}), ordered precursor phase (\textbf{PPo}) and disordered precursor phase (\textbf{PPd}). The corresponding regions in wider range images are marked in (a) and (c) with the same colour code. \textbf{(c)} STM topography images of different samples where the amount of deposited material is the same, whilst the substrate temperature has been increased as indicated by the red arrow on the right. The phases of \ce{FeCl2} that appear in each case are shown. $V_b$ for each case is given in the images. }
\end{figure} 

\ce{FeCl2} is deposited from a commercial Kentax resistive evaporator at a temperature $T_{ev}$ previously calibrated with a quartz micro balance, at a pressure lower than 7$\times10^{-10}$~mbar. The material sublimates for $T_{ev}>$320~$^\circ$C and the evaporation rate increases by one order of magnitude at $T_{ev}$=370~$^\circ$C. The coverage is also very sensitive to the substrate temperature during deposition, $T_s$. We define one ML as a complete coverage of the crystalline 1T phase shown in Fig.~1 of the main text. In the sub-ML regime, there is another competing aggregation phase: a self-assembly of \ce{FeCl2} molecular units stabilized flat on the surface, with both \ce{Cl} atoms in contact with the Au atoms, which can show from long-range ordering to just short range correlation of the intermolecular distances. This is the so-called precursor phase (PP), previously reported for other halides like \ce{CoBr2}~\cite{kerschbaumer_coverage_2025} and \ce{FeBr2}~\cite{hadjadj_febr2_2023}. As shown in Fig.~\ref{fig:FeCl2prep}, the degree of PP ordering, and the prevalence of the PP or 1T bulk phase, is determined by the balance between total dose and $T_{s}$. In its ordered version (blue rectangle), the lattice parameter is indistinguishable by STM or LEED. 

For low $T_s<100$~$^\circ$C and low coverage $<$0.5~ML, the PP is preferred. If the dose is increased close to 1~ML, the material reorganizes into the 1T phase with Fe atoms in octahedral coordination with \ce{Cl} ions (sketch in Fig.~1 of the main text), as shown in the Fig.~\ref{fig:FeCl2prep}, left panel. The right panel illustrates how the 1T phase can also be stabilized at higher temperatures. For instance, for intermediate $T_s>$120~$^\circ$C and small coverages of 0.3 to 0.7~ML, the 1T phase coexists with the ordered PP (PPo). In order to achieve an optimum 0.5 ML coverage of the 1T crystalline phase for this study, $T_s\sim$150~$^\circ$C is required. At higher temperatures $T_s=175$~$^\circ$C, only 1T phase exists for any coverage. The arch shaped LEED spots corresponding to the \ce{FeCl2} suggest that the 1T phase has a good epitaxial registry with the Au(111) lattice directions (see Fig.~\ref{fig:FeCl2prep} central panel), although there is some degree of misalignment. This is in agreement with our atomically resolved images of \ce{FeCl2} 1T slabs, where we have seldom found misalignments higher 8$^\circ$, and a maximum deviation of 15$^\circ$. 

\section{Extended treatment of the interference patterns}

\begin{figure}[h]
\includegraphics[width=0.7\linewidth]{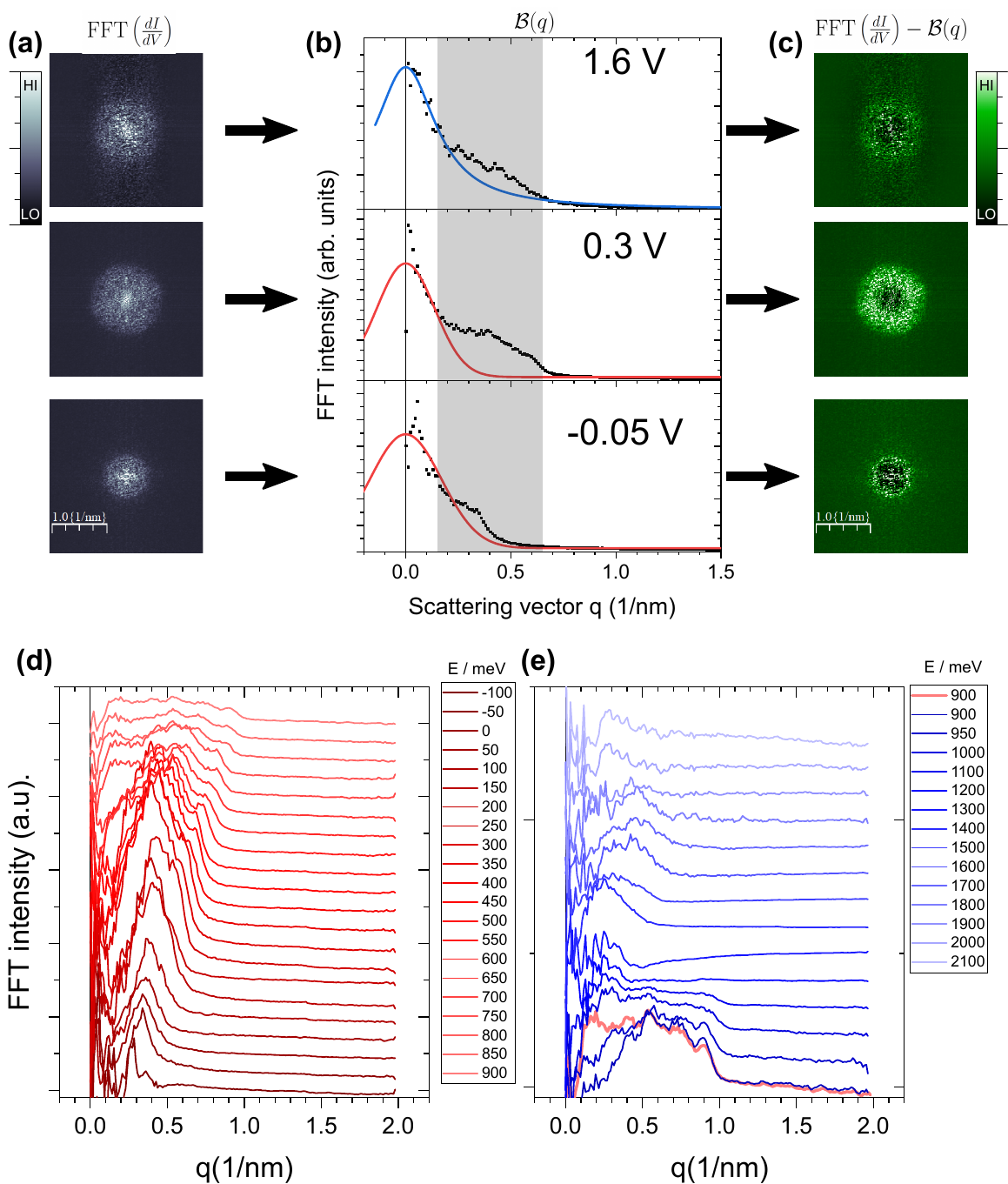}
\caption{\small \textbf{(a)} FFT images of the \ce{FeCl2} region of $dI/dV$ maps taken at, from top to down, $V_b=$1.6~V, 0.3~V and -0.05~V. \textbf{(b)} The corresponding fits to Gaussian (red lines) and Lorentzian (blue lines) backgrounds of the radially averaged images in (a) excluding the shaded area from the fit. \textbf{(c)} Same as in (a) after subtracting the background functions determined as in (b). \textbf{(d,e)} Radial average of the FFT images after background removal for the energies listed in the legend. Red and blue lines represent the use of Gaussian or Lorentzian background functions, respectively.}
\label{fig:FFTmath}
\end{figure}

Au(111)/\ce{FeCl2} FFT images in Fig.~2 of the article are obtained from the same $dI/dV$ map after masking the \ce{FeCl2}/Au(111) region, respectively. To better visualize the energy dependence of the FFT images utilized to compose panels (d) to (f), we have previously subtracted a radial background function $\mathcal{B}(q)$, where $q=(q_x^2+q_y^2)^{1/2}$. $\mathcal{B}(q,E)$ is determined for each energy $E=eV_b$ as a Gaussian ($E\le 0.95$~eV) or Lorentzian ($E\ge 1$~eV) fit to the radial average of the raw FFT image (see Fig.~\ref{fig:FFTmath}a) excluding the $q$ region of interest, as shown by the grey shaded box in Fig.~\ref{fig:FFTmath}b. This background removes the rapidly varying noise of the images (large $q>0.65$~nm$^{-1}$), as well as the long pitch oscillations in real space for which our 50$\times$50~nm$^2$ maps do not provide enough resolution in reciprocal space (small $q<0.15$~nm$^{-1}$). The latter is often contributed by topographic features such as the triangular defects or step edges. The final reciprocal space maps after subtraction of $\mathcal{B}(q,E)$ are shown in Fig.~\ref{fig:FFTmath}c for a few illustrative $E$ values. 

The consistency of this method can be verified by plotting the radial average of the FFT images after background removal, as done in Figs.~\ref{fig:FFTmath}d,e where we show a waterfall representation of the resulting profiles for the case of Gaussian ($E\le0.9$~eV) and Lorentzian ($E\ge0.9$~eV) backgrounds, respectively. The overlap of the FFT radial profile with both methods for $E=0.9$~eV indicates that the functional form of the background is irrelevant to the quantitive analysis in the $q$-range of interest. In both datasets, a clear systematic dependence of the FFT profiles on the energy can be observed. We built the color map in $E-q$ space of Fig.~2~(f) using these profiles after normalizing to their maximum value. The same procedure is applied to the Au(111) FFT patterns to compose Fig.~2~(e).

To extract the effective mass, the surface state of Au(111) and the dispersion of the interface state of \ce{FeCl2} have been fitted with a polynomial of degree 2 over the peak position in $E-q$ space:
\begin{align}
E(q)=C\cdot q^2+B\cdot q+E_0
\end{align}
with $E$ expressed in eV and $[q]=\mathrm{nm}^{-1}$. Then to get the effective mass, $m^*$, from the energy dispersion we use:
\begin{align}
E(k)&=\frac{\hbar^2}{2m^*}\cdot k^{2}+E_0
\end{align}
Since we express $C$ in $\mathrm{eV}\cdot \mathrm{nm}^2$ and $2\pi q=2k$, we have:
\begin{align}
m^*=\frac{0.376}{C}m_e
\end{align}

where $m_e$ is the free electron mass. The coefficients of the best minimum-squared polynomial fit to the Au(111) surface state in Fig.~2~(e) of the main article are: $C=1.657(20)$, $B=0$, $E_0=-0.491$~eV. $E_0$ is fixed to the position of the inflection point in the $dI/dV$ reference spectrum of the Au(111) surface state (red line in Fig.~2~(c)). The effective mass of Au(111) then yields $m^{*} = 0.23m_{e}$, in very good agreement with ref.~\citenum{reinert_direct_2001}. The dashed line describing the interference disk of the \ce{FeCl2} interface state in Fig.~2~(f) is obtained by simply shifting the onset in Eq.~(2) $E_0^\prime=E_0+0.19$~eV, i.e., with the same effective mass as the Au(111) surface state. 
 
Applying the same method (Eqs.~(1-3)) to the parabolic fit of the \ce{FeCl2} conduction band (orange line in Fig.~2~(f)) we obtain  $m^{*} = 1.26m_{e}$ and $E_0=1.54(1)$~eV. \\

\newpage
\section{Spin-polarized measurements}

\begin{figure}[h]
\includegraphics[width=0.9\columnwidth,keepaspectratio]{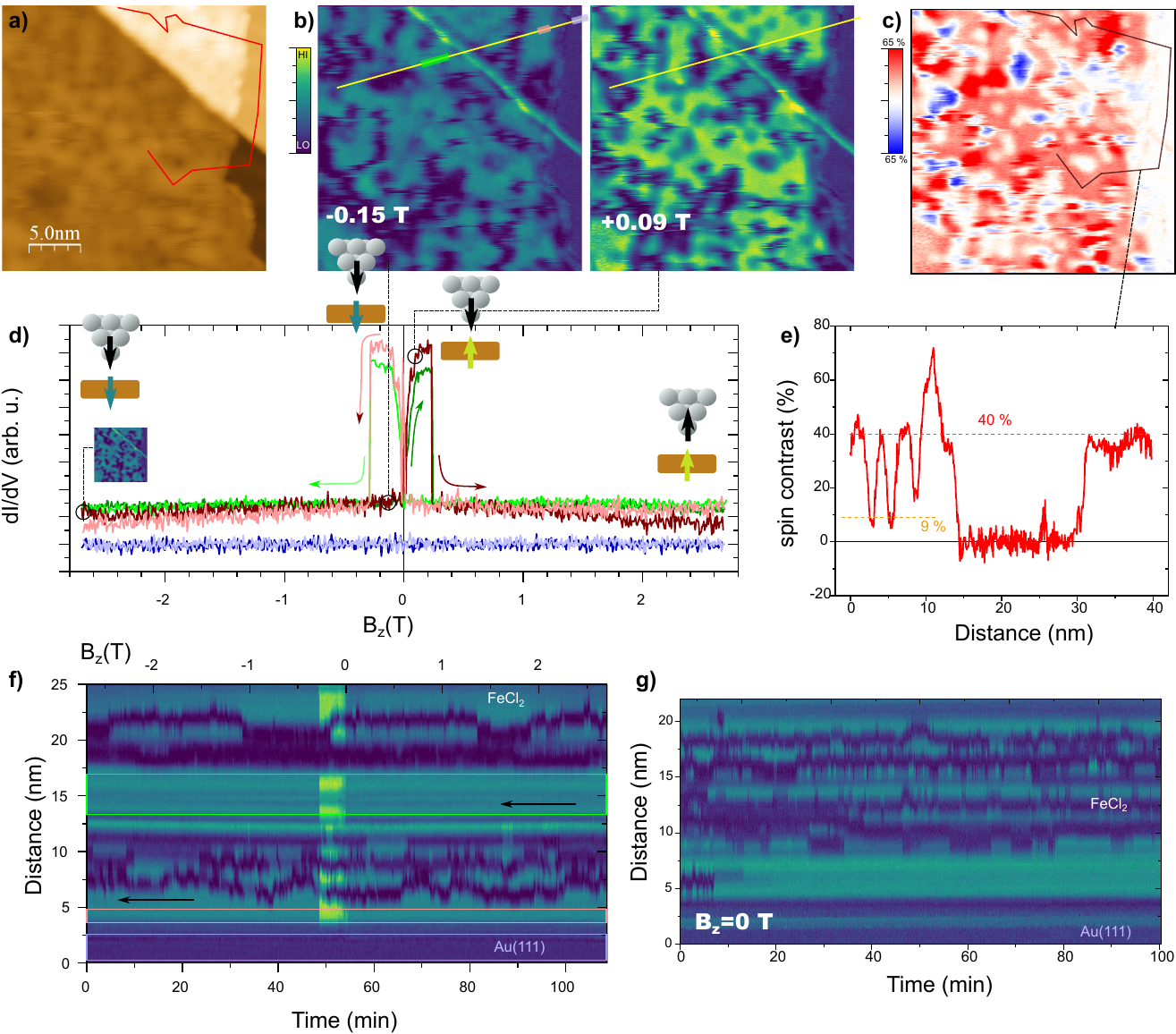}
\caption{\label{fig:maghist} \small \textbf{(a)} STM topography of a \ce{FeCl2} island covering two Au(111) steps. \textbf{(b)} $dI/dV$ maps at $V_b=1.5$~V  of exactly the same area for two opposite orientations of the \ce{FeCl2} magnetization, $dI/dV_{\downarrow\downarrow}$ and $dI/dV_{\uparrow\downarrow}$ at -0.15~T and 0.09~T respectively. \textbf{(c)} Spin contrast map calculated as $\left(dI/dV_{\uparrow\downarrow}-dI/dV_{\downarrow\downarrow}\right)/\left(\langle dI/dV_{\uparrow\downarrow}\rangle +\langle dI/dV_{\downarrow\downarrow}\rangle\right)$, where $\langle dI/dV \rangle$ is the mean differential conductance in the regions without triangular defects. \textbf{(d)} $dI/dV$ intensity at $V_b=1.5$~V as a function of axial magnetic field taken by continuously scanning the yellow line shown in panel (b), and extracted independently for the two different terraces of \ce{FeCl2} (at each field value, we take the average $dI/dV$ across the green and reddish segments on top the yellow line) and the Au(111) region (blue segment). Thin coloured arrows depict the field variation direction for each one of the traces with the same colour code. Dashed lines mark the field loop status at which $dI/dV$ maps in (b) were acquired, and the cartoons depict the corresponding tip and sample magnetic states. The small inset shows the $dI/dV$ map at -2.7~T. \textbf{(e)} Spin contrast profile along the line shown in panels (a) and (c). \textbf{(f)} Map of stacked line profiles along the yellow line of panel (b) as a function of axial field and time. Rectangles mark the areas in which the field dependent traces of panel (d) with the same colour code are averaged. This data set corresponds to the field descent part of the loop. \textbf{(g)} Comparison of the same type of line scan as in panel (f) in a different island at constant zero field, showing that the $dI/dV$ fluctuations caused by the jumps of the defects are not driven by the magnetic field. Panels (b), (f) and (g) share the same colour scale. STM parameters $V_s=1.5$~V, $I_s=$100~pA. Lock-in modulation 10~mV rms. $T=1.1$~K.}
\end{figure}

We report in Figs.~3 and 4 of the article spin resolved measurements of the conduction band onset observed at $V_b=1.5$~V. Figure~\ref{fig:maghist} shows the same data set for a larger area and magnetic field range. In the spin contrast map of Fig.~\ref{fig:maghist}~(c,e) there are two kind of intensity levels. One is the spin density of the areas free of triangular defects (TD) with a magnetoconductance contrast of about 40\%, which vanishes completely in the non-magnetic Au regions. This is the intrinsic contrast of the 1T \ce{FeCl2} ferromagnetic phase. 

There is, however, another stronger contrast which reaches maximum conductance variations of $\pm$70\% between $dI/dV$ images taken at different fields. This is an artifact caused by the fluctuation of TD occurring in between the acquisition of both images. At positions where a TD appears in the last image, the disappearance of the CB onset (see main article) induces a negative (intense blue) conductance variation of around -70\%. At regions where a TD drifts away to another position, the CB onset causes the opposite conductance variation of about +70\% (intense red). In Figs.~\ref{fig:maghist}~(f), (g) we show the time dependence of such TD fluctuations. The presence of a TD manifests as a drop in the local $dI/dV$ signal at $V_b=1.5$~V, which jumps back and forth around the original position at slow attempt rates between 0.1-0.001~s$^{-1}$ (the largest the jump, the longer the lifetime of a TD on a particular position). The attempt rate also increases for increasing regulation $I_s$. The quantitative analysis of the spin contrast was performed in regions that were not visited by any TD during the measurements, as is the case of the fragments enclosed by coloured rectangles in Fig.~\ref{fig:maghist}~(f). It is to be noted that the TD fluctuation is not related with the magnetic field variation, as can be deduced by comparing Figs.~\ref{fig:maghist}~(f) and \ref{fig:maghist}~(g), taken with varying and constant magnetic fields, respectively, during comparable time periods.\\

\clearpage




\twocolumngrid



\end{document}